\documentclass[prb, preprint,showpacs]{revtex4}
\usepackage{graphicx}
\begin{document}
\title{Peak resistance temperature and low temperature resistivity in thin film 
La$_{1-x}$Ca$_{x}$MnO$_{3}$ mixtures for $x\ \leq$ 1/3} 
\author{P. R. Broussard} 
\affiliation{Covenant College, Lookout Mountain, GA 30750} 

\begin{abstract}
The electrical resistivity of La$_{1-x}$Ca$_{x}$MnO$_{3}$ thin films grown on (001) 
NdGaO$_{3}$ and (100) SrTiO$_{3}$
substrates by off-axis sputtering has been studied as a function of the Calcium doping level.  
The samples have very narrow rocking curves and excellent in plane 
registry with the substrate.  A strong correlation between the peak resistance temperature and the polaronic hopping energy is seen which is not simply linear.  The low temperature resistivity is seen to be fit better by a model of single magnon scattering, and a near linear correlation between the resistivity due to magnon scattering and static impurities is observed.
\end{abstract}
\maketitle

\section{Introduction}
The discovery of colossal magnetoresistance (CMR) in the doped manganite materials\cite{Helmolt} and subsequent study of the properties of these materials has created a great deal of interest in both the application and the underlying cause of CMR.  The connection between the Double Exchange mechanism and the Jahn-Teller lattice distortions has focused attention on the carriers in the systems.  Much work has been done to show that the adiabatic small polaronic model fits the resistivity above the metal-insulator (MI) transition in the La$_{1-x}$Ca$_{x}$MnO$_{3}$ (LCMO) system.\cite{Worledge, Song, Hartinger}  
In an earlier publication,\cite{PRB1} the anomalous 
behavior of 
the peak resistance temperature for La$_{1-x}$Ca$_{x}$MnO$_{3}$ films 
with $x$ varying from 0 to 1/3 was mentioned, where the temperature for the metal-insulator transition ($T_{MI}$) or the peak resistance temperature ($T_{p}$) was higher than that seen in bulk materials for $x< 0.25$ and that the metal insulator transition persisted below the $x=0.175$ value which is the boundary in bulk materials for this transition.  Similar results were seen in the work by Prellier {\it et al.}\cite{Prellier} for films of La$_{1-x}$Ca$_{x}$MnO$_{3}$ for $0.1<x <0.5$ grown on LaAlO$_{3}$.  Recent work on LCMO material has addressed the connection between the polaronic hopping energy and the peak resistance temperature\cite{Song}, as well as the behavior of the low temperature resistivity in the magnanites.\cite{Mercone} In this paper, the films studied in the earlier work\cite{PRB1} are examined in light of these recent analysis.  In particular, it is shown that 1) there is a definite relationship between the polaronic hopping energy and the peak resistance temperature, but the relationship is more complicated than Song {\it et al.} found, and 2) the low temperature behavior of the resistivity is more consistent with the model by Jaime {\it et al.}\cite{Jaime1}

\section{Sample preparation and characterization}
The samples were grown by off-axis sputtering using composite targets
of La$_{0.67}$Ca$_{0.33}$MnO$_{3}$ (LCMO) and LaMnO$_{3}$ (LMO)
material mounted in copper cups.  The substrates were (001) oriented
neodymium gallate (NdGaO$_{3}$) and (100) oriented strontium titanate
(SrTiO$_{3}$), silver-pasted onto a stainless steel substrate holder
that was radiatively heated from behind by quartz lamps.  Although
there was no direct measurement of the holder temperature for the runs
used in this study, previous runs (under nominally the same
conditions) using a thermocouple clamped onto the front surface of the
holder indicated a temperature of 670 $^{\circ}$C. The LMO target was radio
frequency sputtered and the LCMO target was direct current
sputtered in a sputter gas composed of 80\% Ar and 20\% O$_{2}$ (as
measured by flow meters) and at a total pressure of 13.3 Pa.  These
conditions gave deposition rates of $\approx$ 17-50 nm/hr, with film
thicknesses being typically 150-300 nm.  After deposition, the samples
were cooled in 13.3 kPa of oxygen.  From previous work\cite{PRB2}
it is known that this process can grow manganite films that have low
resistivities and high peak temperatures without the use of an {\it
ex-situ} anneal in oxygen.  

The Ca concentration in the films, $x$, was determined by X-ray fluorescence
measurements, which in turn were calibrated by Rutherford
Backscattering (RBS) for companion samples deposited on MgO
substrates.  The samples were characterized by standard and high
resolution x-ray diffraction scans (using a Philips MRD 
x-ray diffraction system, with a four-crystal Ge 220 monochromator on the 
incident beam and CuK$\alpha_{1}$ radiation ($\lambda$=1.5045 $\AA$)).  $\theta-2\theta$ scans 
were taken both along and at various angles to the growth direction, as 
well as rocking curves and $\phi$ scans (where $\theta-2\theta$ is set 
for a particular reflection at an angle $\psi$ to the film normal, and 
the film is rotated about the film normal) to look for misoriented grains.  
Electrical resistivity measurements were made using the van der
Pauw method\cite{VdP}.

\section{Structure of the films}

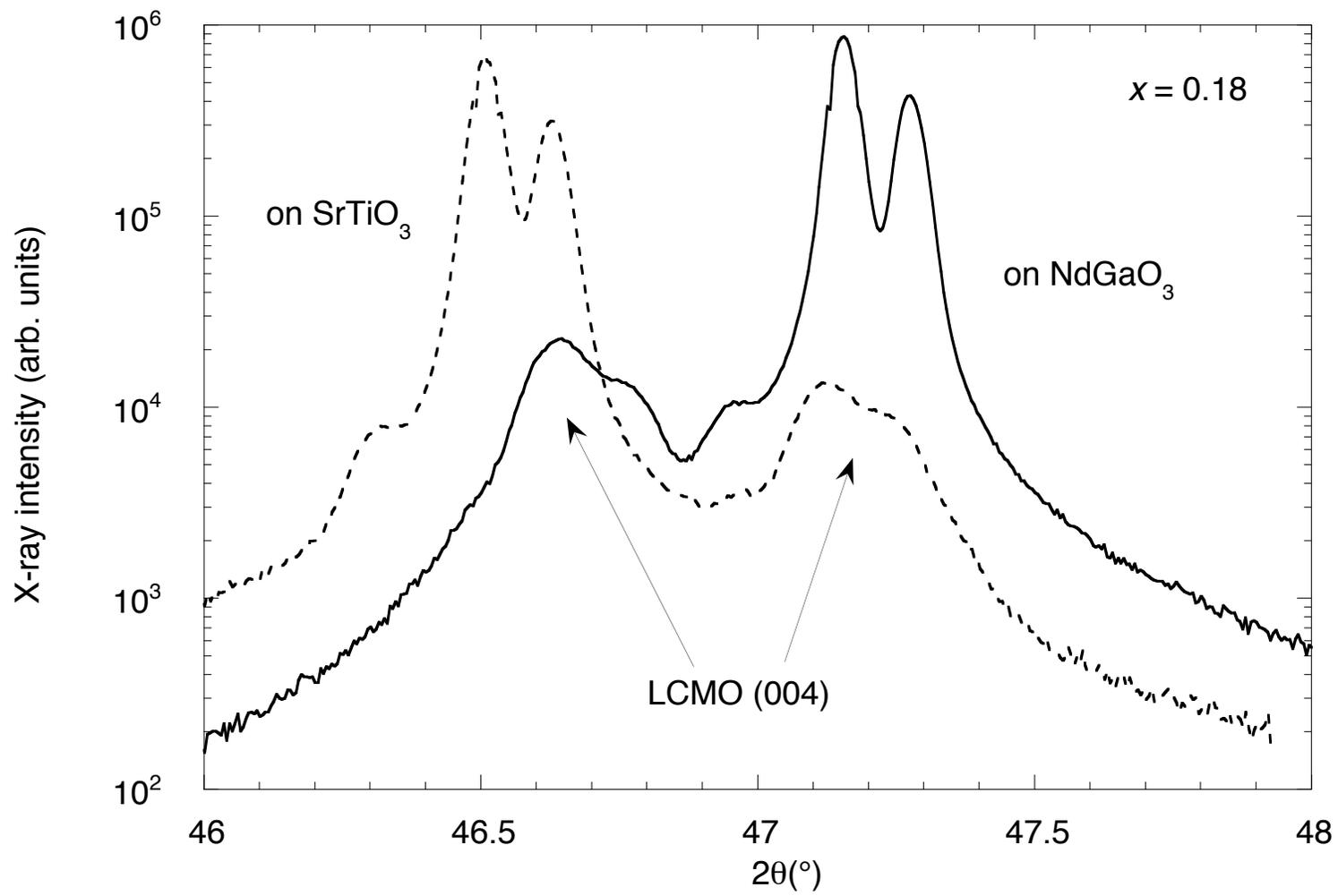
\begin{figure}
\begin{center}
\end{center}
\caption{X-ray diffraction scans along the film normal 
for LCMO film ($x$=0.18) grown on SrTiO$_{3}$ (dashed) and NdGaO$_{3}$ (solid) 
substrates.}
\label{XRD1}
\end{figure}

X-ray diffraction shows that the films on both substrates are oriented (00L) 
(orthorhombic notation) perpendicular to the
film plane.  Fig. \ref{XRD1} compares the x-ray diffraction scans taken in 
standard resolution mode (Cu K$\alpha$ radiation) for a film with $x$=0.18 on SrTiO$_{3}$ and 
NdGaO$_{3}$.  The reduction in the c-axis lattice 
constant induced by the film attempting to lattice match to the 
larger lattice constant of SrTiO$_{3}$ and the opposite for the case 
on NdGaO$_{3}$ is clearly seen.  The c-axis lattice constant increases as the Ca 
concentration is decreased, similar to that seen in bulk material.\cite{Huang} The samples on NdGaO$_{3}$ show more strain 
for low values of $x$ while the samples on SrTiO$_{3}$ have more strain near $x\approx 1/3$.  The samples on NdGaO$_{3}$ are on average more strained than samples on 
SrTiO$_{3}$ are for intermediate values of $x$.

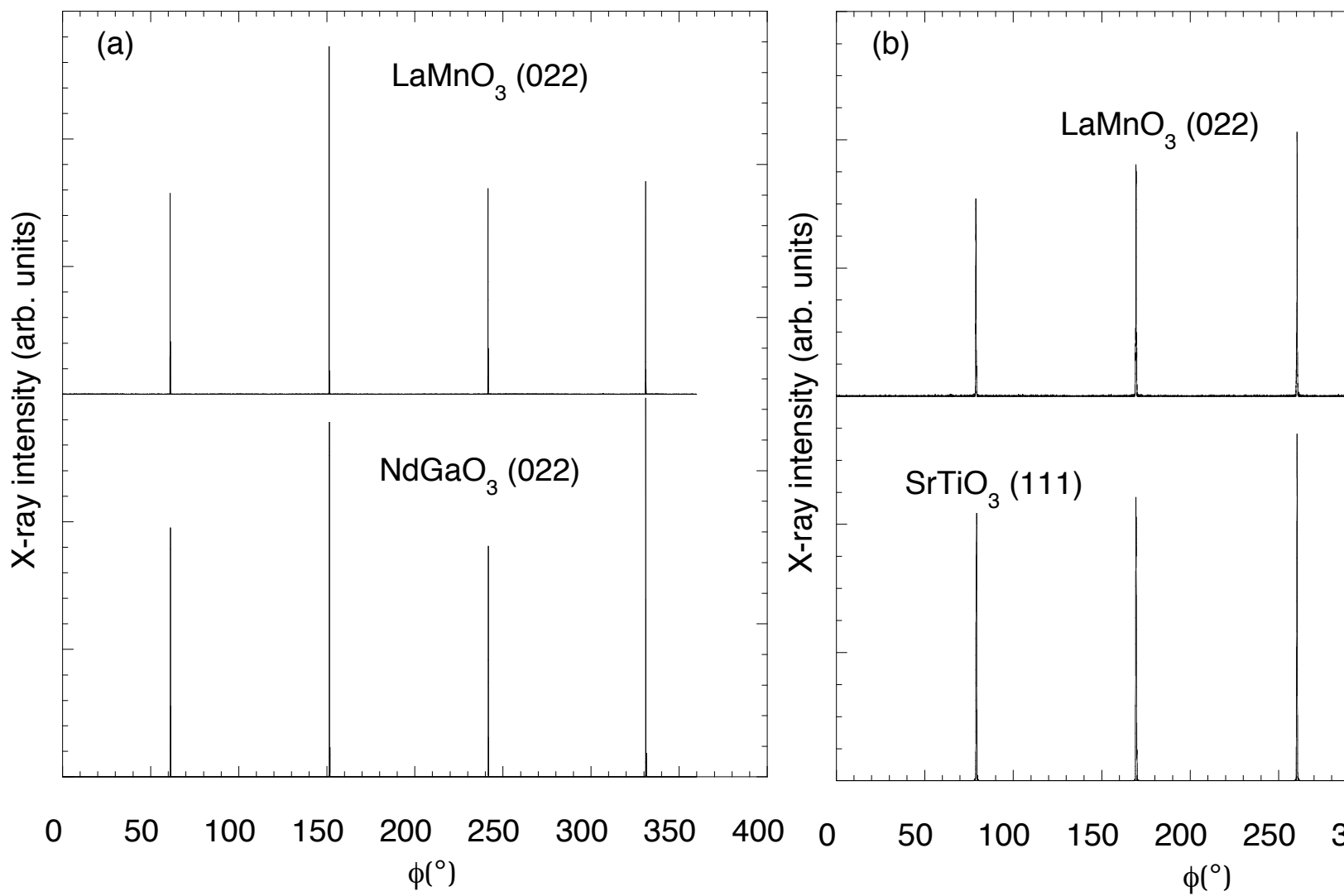
\begin{figure}
\begin{center}
\end{center}
\caption{X-ray diffraction phi scans (a) along the LMO (022) (top) and 
NdGaO$_{3}$ (022) lines 
for LMO film ($x=0$) grown on NdGaO$_{3}$ and (b) along the LCMO (022) and SrTiO$_{3}$ (111) for a LCMO film ($x = 1/3$) grown on SrTiO$_{3}$.  The curves are offset for 
clarity.}
\label{XRDphi}
\end{figure}

The film quality for the samples on NdGaO$_{3}$ is excellent, with 
rocking curve widths for the (004) reflection of the order of 200 arc -seconds (with an instrumental resolution of 15 arc-seconds).  The 
films on SrTiO$_{3}$ had broader rocking curve widths, typically of 
order 300 arc-seconds, partly due to the fact that the SrTiO$_{3}$ 
substrates have large rocking curve widths since they are grown by a flame fusion process, rather than 
the Czochralski method used for NdGaO$_{3}$ substrates.  The in-plane order 
for the films, however was similar between the two substrates. 
Fig. \ref{XRDphi} shows a phi-scan for the (022) reflections for 
both NdGaO$_{3}$ and LaMnO$_{3}$ for the $x=0$ sample grown on NdGaO$_{3}$ and a similar plot for the $x=1/3$ sample grown on SrTiO$_{3}$.  Excellent 
registry between the film and the substrate is seen in both cases.  (The reason for the 4-fold 
rather than the expected 2-fold symmetry for both materials is due to 
the acceptance angle on the detector seeing both the (202) and (022) 
reflections.)  The rocking curve widths for the (022) LCMO reflections 
are similar to those seen for the (004) reflections.

\section{Peak temperature and resistivity}

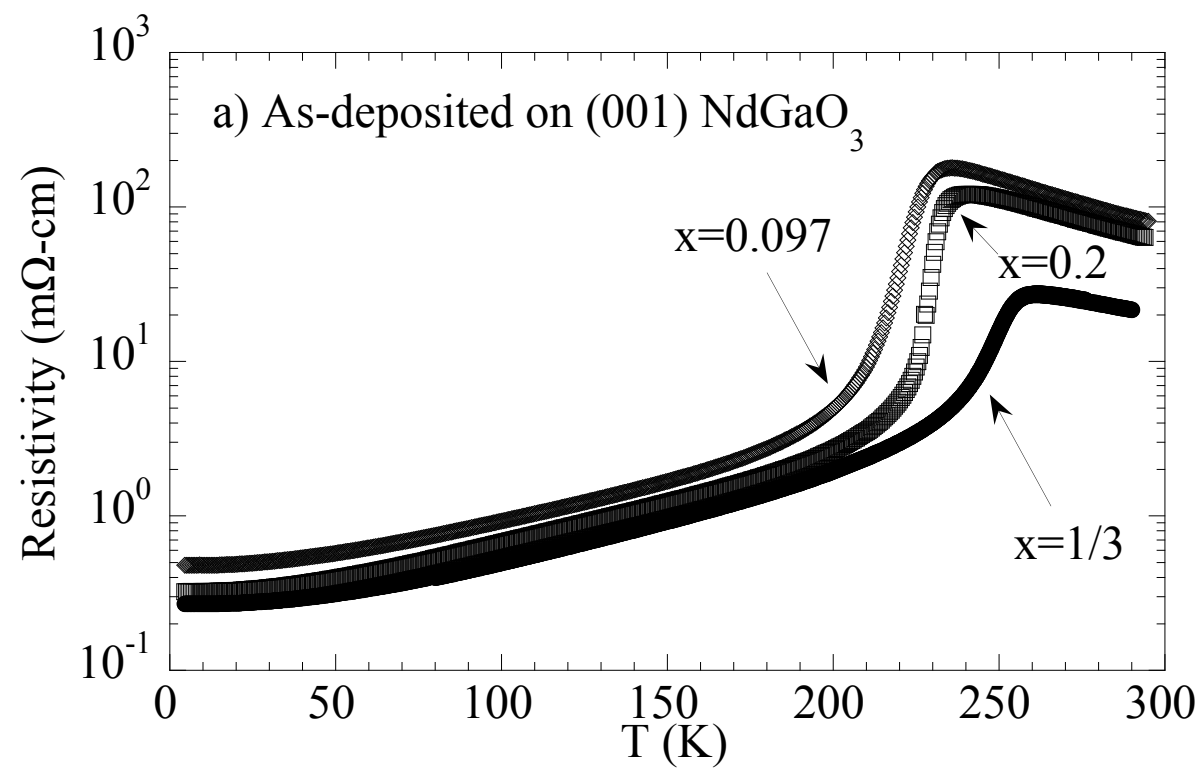
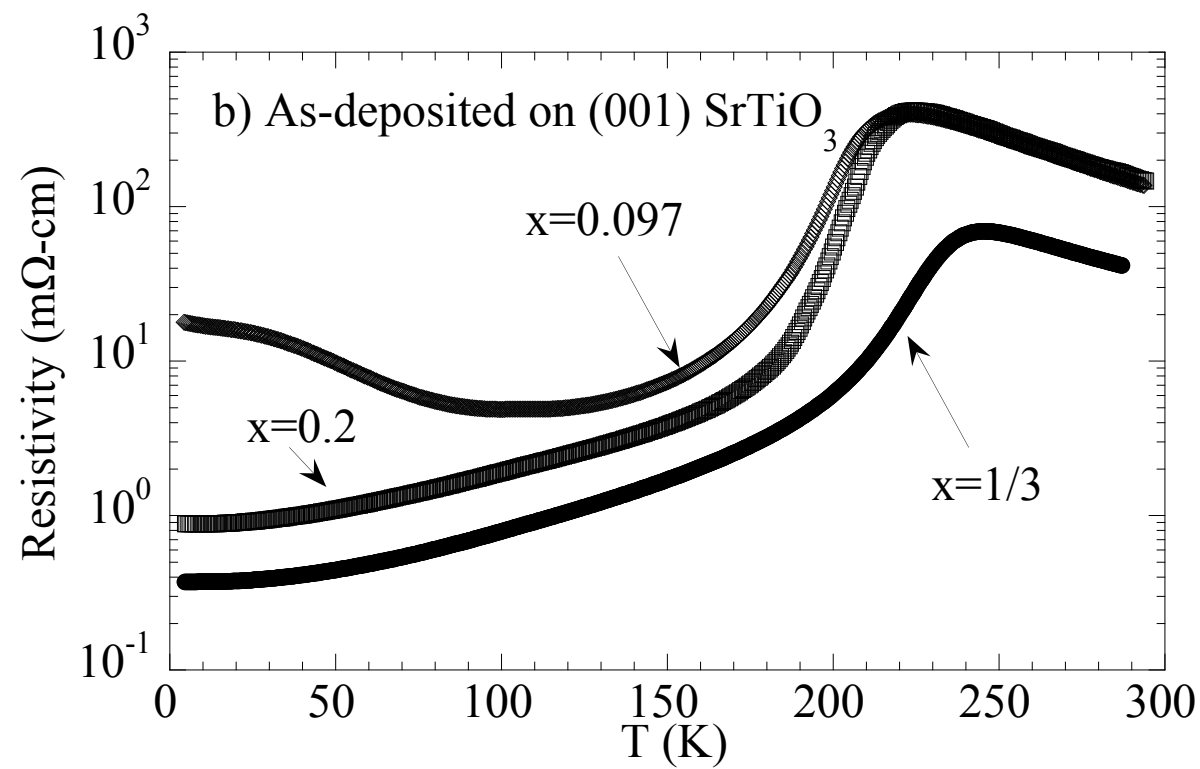
\begin{figure}
\begin{center}
\end{center}
\caption{Resistivity vs. temperature for as deposited 
La$_{1-x}$Ca$_{x}$MnO$_{3}$ films on 
a) (100) NdGaO$_{3}$, and b) (001) SrTiO$_{3}$ for various values of $x$.}
\label{Rho1}
\end{figure}

Fig. \ref{Rho1} shows the resistivity vs. temperature for 
several of the 
as-deposited films, which have low resistivities and high peak transition
temperatures ($T_{p}$).  Typically, the films on SrTiO$_{3}$
have values of $T_{p}$ that are $\approx$ 15 K lower than films on NdGaO$_{3}$, as well as
higher resistivities, even though the films on NdGaO$_{3}$ are under more strain usually.    For samples with $x=1/3$, the resistivity at low temperatures is less than 0.4 m$\Omega$-cm, and show a greater than two order of magnitude reduction from the resistivity at $T_{p}$, both of which have been used as measures of film quality.\cite{Song, Mercone}  As was noted in the earlier work\cite{PRB1}, the samples on both substrates maintain a high value of $T_{p}$ even for low values of $x$.  This report now turns to two areas of the resistivity of the samples. 

\subsection{$T_{p}$ and high temperature resisitivity}

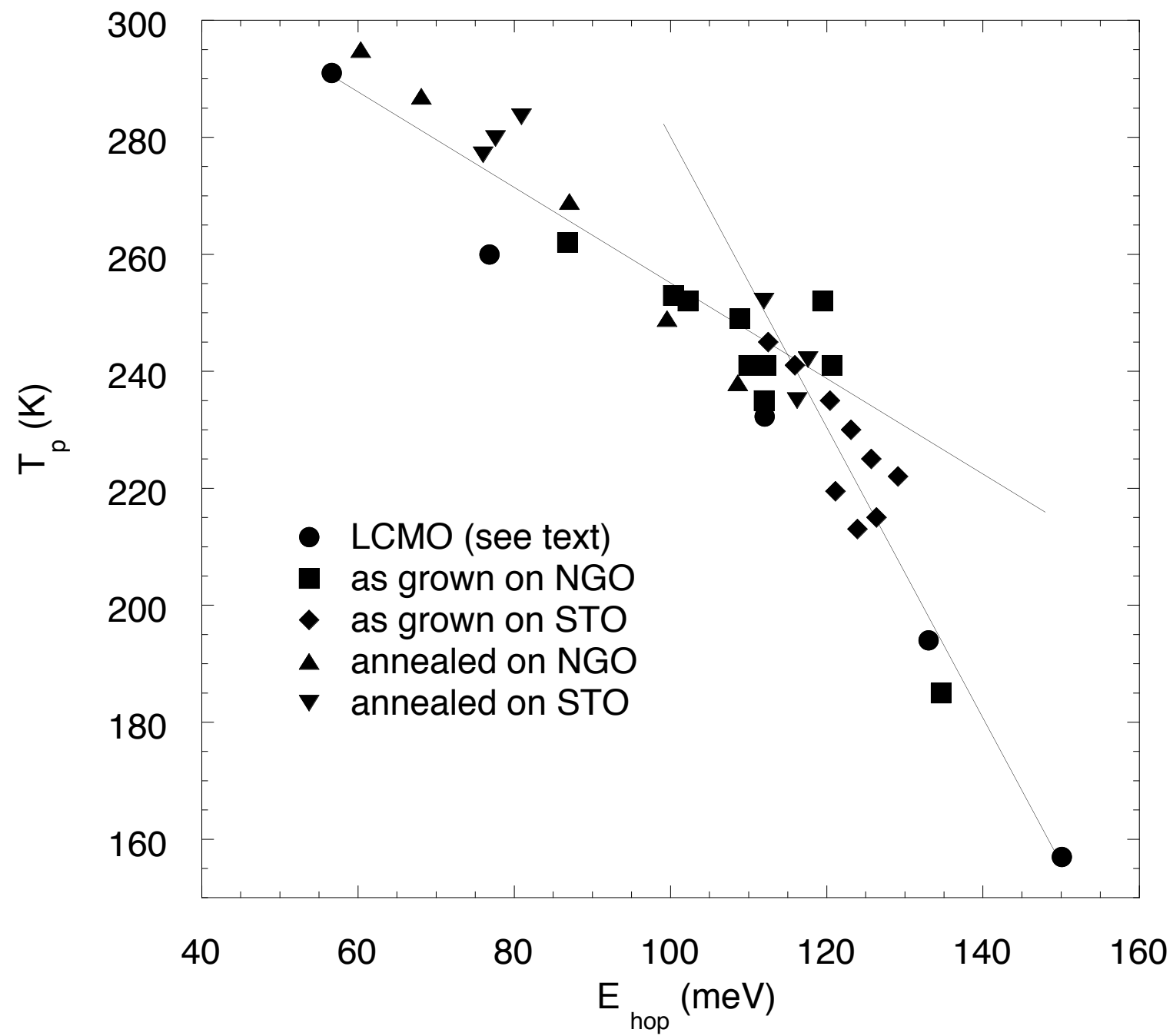
\begin{figure}
\begin{center}
\end{center}
\caption{Peak resistance temperature vs. polaronic activation energy for as grown and annealed LCMO films on NdGaO$_{3}$ (NGO) and SrTiO$_{3}$ (STO).  The films labeled LCMO are from other samples of LCMO grown on a variety of substrates and different growth conditions by the author.  The lines are guides to the eye.}
\label{TpEA}
\end{figure}

Much work has been done to understand the interaction between the Double Exchange model and the CMR effect in these materials.  Presently the connection is made that polaron interactions are responsible for the CMR effect, which implies that above the peak temperature the resitivity should be dominated by the adiabatic small polaronic resistivity, which is given by $\rho(T)\propto T\exp{\frac{E_{hop}}{k_{B}T}}$.  For the temperature range considered in this study, both the small polaron model as well as the Arrhenius relation ($\rho(T)\propto\exp{\frac{\Delta_{p}}{k_{B}T}}$) fit the data equally well.  In the work by Song {\it et al.}\cite{Song}, they found a correlation between the activation energy, $E_{hop}$, and the value of $T_{p}$ for oxygen deficient LCMO films, which they intepreted as a linear dependence.  For the films in this study, Fig. \ref{TpEA} presents a similar study, including results for films of this study that were subsequently annealed to 700 $^{\circ}$C in oxygen, which resulted in an increase in their peak temperature.    Clearly in this case the dependence between $T_{p}$ and $E_{hop}$ is not linear over the whole range, although there is direct relationship between the two.  This dependence is very similar to what was observed by Browning {\it et al.}\cite{Browning} where they found a correlation between the activation energy (assuming a simple Arrhenius relation) and the Curie temperature of radiation damaged LCMO films.  One can see that over a limited range of this plot, a linear relationship could be found, but clearly the whole range is more complicated.  In fact the plot seems to show that as the polaronic hopping energy gets large enough, the metal-insulator transition is more rapidly depressed.  What is interesting is that the values here are similar to those found by Song {\it et al.} even though their values are for a fixed cation composition and varying oxygen contents, while the films here are for varying dopant levels.  The plot also presents results for LCMO films produced in a similar manner as the films here but on different substrates or growth temperatures.  These are labeled as ``LCMO'' on the plot.  Clearly there seems to be a behavior that is more universal than just for carefully controlled samples.

\subsection{Low temperature resisitivity}

Several attempts have been made to describe the low temperature behavior of
the resistivity for the manganate materials.  In an attempt to explain data
which implies a $T^{2.5}$ dependence, such as seen in Schiffer et
al.\cite{Schiffer}, Wang and Zhang\cite{WZ} have looked at the system with
single magnon scattering and where the minority spin states near the Fermi
edge are Anderson localized.  They predict that the resisitivity will go as
$T^{2.5}$ until the temperature decreases below $\approx$ 60K, when the
behavior changes to $T^{1.5}$.  Recently, Mercone {\it et al.}\cite{Mercone} saw the resisitivity of LCMO and LSMO films grown by various techniques behaving as $T^{\alpha}$ with $\alpha\approx 2.5$ for the low temperature range.  Jaime {\it et al.}\cite{Jaime1} have also looked
at the case of single magnon scattering, but in their case the minority
carriers are not localized.  In this model, the single magnon scattering
term will go as $T^{2}$, but with a temperature dependent coefficent that
is due to an energy shift between the minority and majority spin bands. 
The signature of this type of behavior is a strong reduction in the $T^{2}$
term as the temperature is decreased below the minimum magnon energy that
would allow scattering between the spin up and down energy bands.

As observed in earlier 
work\cite{PRB2} and  is observed here, the low temperature 
resisitivity for these films is better fit by
\begin{equation}
\rho (T) = \rho_{0} + \rho_{2}T^{2} +\rho_{5}T^{5}.
\label{rhoT}
\end{equation}
The present data was examined to see if it can be explained by the work of 
Wang and Zhang, especially with the predicted change in temperature 
dependence, to no avail.  What is found is that, as was seen in 
Jaime {\it et al.}, there is a reduction in the $T^{2}$ term, typically 
below 30K.  As such, the low temperature resistivity data for 
the films was fit using Eq.\ref{rhoT} modified by the model of Jaime {\it et al.}  The equation used is
\begin{equation}
\rho(T)=\rho_{0}+\frac{\rho_{2}}{\pi^{2}/6}T^{2}\int_{T_{q}/T}^{\infty}\frac{x^{2}}{sinh^{2}(x)}dx +\rho_{5}T^{5}
\label{rhofit}
\end{equation}
where $T_{q}$ represents the term $Dq^{2}_{min}/2k_{B}$ in Jaime {\it et 
al.}, with $Dq^{2}_{min}$ being the minimum magnon energy that would 
allow scattering between the spin up and down energy bands.  
$\rho_{2}$ represents the single magnon scattering contribution to 
the resistivity in the absence of any energy gap.  The fit 
also incorporates a static impurity term, $\rho_{0}$, and an electron-phonon 
term, $\rho_{5}$.  
The data for $T < T_{p}/2$ was fit using the simplex minimization 
routine of Nelder and Mead,\cite{NelMed}  which minimizes an average weighted chi squared,
\begin{equation}	
	\chi^{2}_{W}/N=\frac{1}{N}\sum_{i=1}^{N}\frac{(\rho_{fit}(i)-\rho_{exp}(i))^{2}}{\rho_{exp}^{2}(i)}
\end{equation}
where N is the number of fit points (typically 400).  
The program would rerun the 
minimization 5 times from random locations in $(\rho_{0}, \rho_{2}, 
T_{q},$ and $\rho_{5})$ space to ensure the ``best'' minimization had 
been found.  Typical values of $\sqrt{\chi^{2}_{W}/N}$ (which is a measure of the rms error per point for the fit) were 2-3 x 10$^{-3}$.  The lowest temperature used in the fitting was down to our lowest measured value (6 K or lower), unlike the work by Mercone {\it et al.}, who had to limit to 20 K and higher due to upturns in the resistivity.  Attempts to use a dependence like $\rho(T)=\rho_{0}+AT^{\alpha}$ for the films gave values of $\sqrt{\chi^{2}_{W}/N}$ larger by a factor of 10 than those of Eq. \ref{rhofit}.  The results of this fitting are shown in Tables \ref{rhofitdata1} and \ref{rhofitdata2}.  The only films that were not fit where the ones on SrTiO$_{3}$ which showed resistivity upturns at low temperatures, which happened for $x<0.2$.

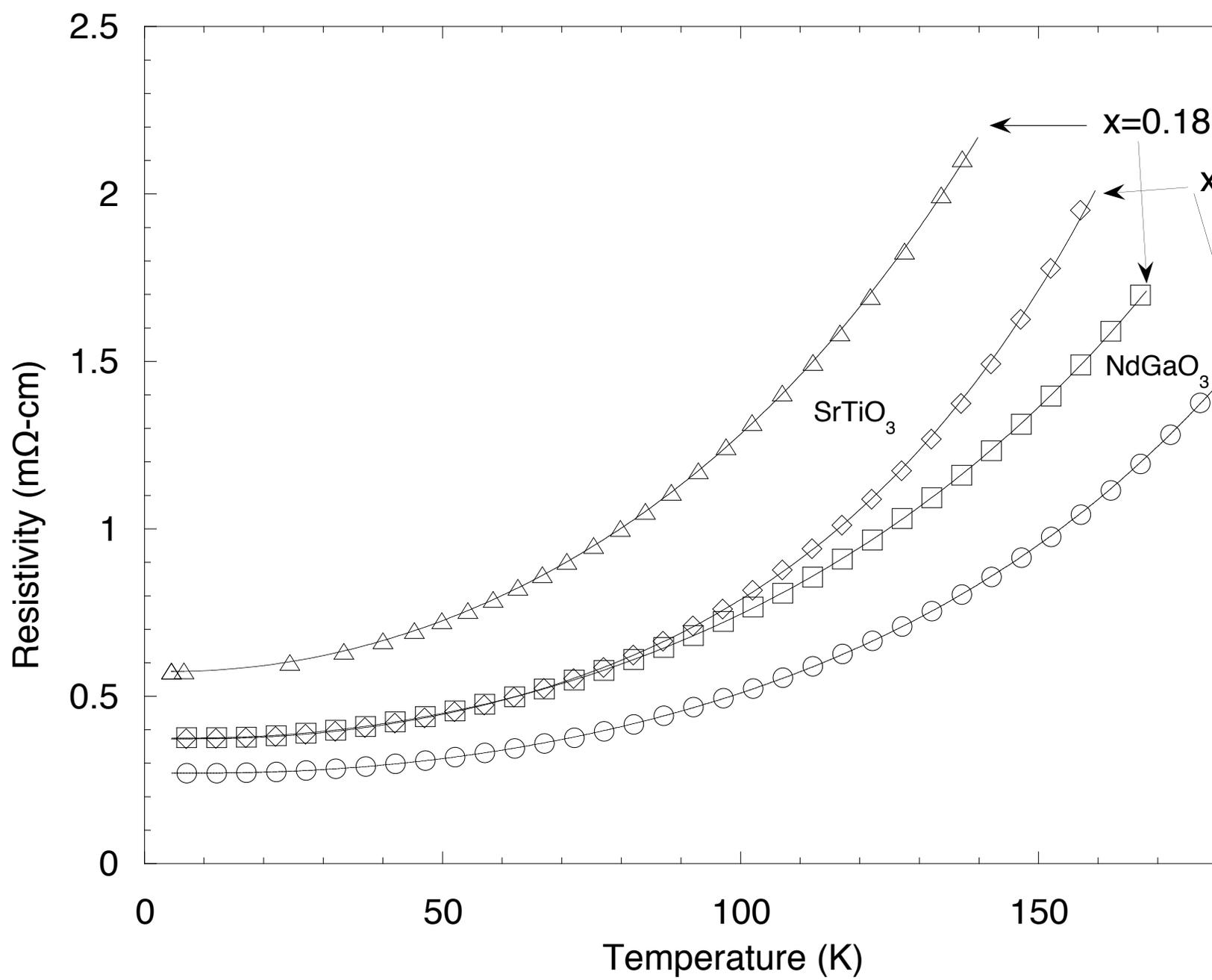
\begin{figure}
\begin{center}
\end{center}
\caption{Low temperature resistivity vs. temperature for the $x=0.18$ 
and 0.33 samples on NdGaO$_{3}$ and SrTiO$_{3}$ substrates along with 
the fits derived from Eq. \ref{rhofit}}
\label{lowrho1}
\end{figure}

In Fig. \ref{lowrho1} the results of fitting Eq. \ref{rhofit} 
to the experimental data for the $x=0.18$ and $x=1/3$ samples on both 
SrTiO$_{3}$ and NdGaO$_{3}$ substrates are shown.  The fit quality is seen to be
very good over the entire temperature range.  To compare between the 
proposed models to describe the low temperature behavior, in Fig. 
\ref{lowrho2} the T $<$ 50 K data for the $x=0.2$ sample on NdGaO$_{3}$ is shown, 
along with the results of fitting Eq. \ref{rhofit}, along with results 
of fitting Eq. \ref{rhoT} as well as the $T^{\alpha}$ behavior seen by Mercone {\it et al.}  One can 
clearly see the $T^{\alpha}$ model cannot reproduce the lowest 
temperature behavior.  One can also see 
that at the lowest temperature, the fit from Eq. \ref{rhoT} 
underestimates the data, which is due to the supression of the 
$T^{2}$ term predicted by Jaime {\it et al.}  

For the films studied here, we see that, in general, there is an increase in $\rho_{0}$ and $\rho_{2}$ and a decrease in $T_{q}$ as the Ca concentration decreases.  The magnitude of $\rho_{5}$ is roughly constant as $x$ is varied, with the all the values being higher for films on SrTiO$_{3}$ (except for $T_{q}$). The value of $\rho_{5}$ found here is in the same range, but higher, than found by Jaime {\it et al.}\cite{Jaime1}, who saw a value around 1 f$\Omega$-cm/K$^{5}$.   Our values of all our quantities are larger than seen in the LSMO/LCMO composite mixtures\cite{PRB2} that were derived using Eq. \ref{rhoT}.  The increase in $\rho_{0}$ as $x$ decreases could be tied to the increase in strain for the samples on NdGaO$_{3}$, but that would not hold for the films on SrTiO$_{3}$, which showed a decrease in strain as $x$ decreases.  In addition, the fact that the values for $\rho_{0}$ being higher on films grown on SrTiO$_{3}$ (which had overall lower strain than the films on NdGaO$_{3}$) would imply that strain is not the dominant contributor to $\rho_{0}$.  The increased rocking curve width for samples grown on SrTiO$_{3}$ would imply an increased static disorder due to grain misalignment, which could explain the higher values of $\rho_{0}$.  Since the spin wave stiffness, $D$, has been seen to be a consistent value across different materials with various doping levels ($\sim$ 170 meV-$\AA^{2}$),\cite{Lynn,Baca} a decrease in $T_{q}$ would imply a reduction in $q_{min}$, the shift in the Fermi momentum between the minority and majority spin bands as the doping decreases. From Jaime {\it et al.} the $\rho_{2}$ term is given by
\begin{equation}
\rho_{2}=\frac{3\pi^{5}(NJ)^{2}\hbar^{5}}{16e^{2}E_{F}^{4}k_{F}}\big(\frac{k_{B}}{m^{*}D}\big)^{2}
\label{rho2eq}
\end{equation}
where $NJ$ is the electron-magnon coupling energy and $E_{F}$, $k_{F}$, and $m^{*}$ are the Fermi energy, Fermi wavevector, and effective mass for the carriers.  It is expected that as $x$ decreases, there will be a reduction in both the values of Fermi wavevector and energy, which would tend to increase the value of $\rho_{2}$.  $D$ might be considered to be constant as discussed above.  However, without knowing how $NJ$ and $m^{*}$ vary with $x$, it will be difficult to draw any firm conclusions.

In the earlier work on LSMO/LCMO composite mixtures,\cite{PRB2} a correlation was seen between the magnitude of the $T^{2}$ coefficient and the residual resistivity in the films.  In this work the same correlation is observed, as shown in Fig. \ref{r2r0}.  This plot shows the relationship between $\rho_{2}$ and $\rho_{0}$ for the films of this study, along with the films from work on LCMO/LSMO composites\cite{PRB2}, as well as the results reported by Jaime {\it et al.}\cite{Jaime1} on La$_{0.66}$(Pb$_{0.67}$Ca$_{0.33}$)$_{0.34}$MnO$_{3}$ single crystals as well as work by Snyder {\it et al.}\cite{Snyder} on LCMO films.  There is a clear linear correlation between these two values, as was seen before.  In the work on LCMO/LSMO composites, the limited data gave an approximate ratio of 60-70x10$^{-6}$ K$^{-2}$, while with the more extensive data here the ratio between the two is more of the order of $\approx$ 120x10$^{-6}$ K$^{-2}$.  The interesting thing is that this correlation is seen not only for LCMO films of varying doping, but is also present in mixtures of LCMO and LSMO as well as crystals of Pb doped LCMO.   The question arises as to the possible connection between the static resistivity and the amount of single magnon scattering.   The work by Snyder {\it et al.} showed a similar behavior for their films, but the point of this paper is that this correlation holds over a wide range of films as well as values of $\rho_{0}$.  For the films of this study, the variation is both due to changes in doping level as well as differering amounts of strain in the samples, which decreases for samples on NdGaO$_{3}$ and increases for those on SrTiO$_{3}$ as the doping increases.  As discussed above, in the model of Jaime {\it et al.} the $\rho_{2}$ term depends on the carrier concentration in the materials, as well as the electron-magnon coupling, the effective mass of the carriers, and the spin wave stiffness.  It is possible to imagine that if strain in the films is one cause of increasing $\rho_{0}$, then there could also be a connection to the scattering of the carriers off spin waves as strain could impact other properties in the expression for $\rho_{2}$.  However, it does not appear that strain in the films is the major cause of $\rho_{0}$, and it would be hard to imagine how say an increase in lattice defects would necessarily increase the effectiveness of magnon scattering.  The question remains as to whether this correlation is wider than presented here as well as what could be causing the connection between these two scattering processes.

\section{Conclusion}

This work has shown that films of La$_{1-x}$Ca$_{x}$MnO$_{3}$ grown on NdGaO$_{3}$ and SrTiO$_{3}$ have excellent structural properties, high values of peak resistance temperatures, and low residual resistivities.  As seen in the work by Song {\it et al.} there is a correlation between the peak resistance temperature and the polaronic hopping energy, but the correlation is not simply linear, but is more complicated, with a more pronounced reduction in the value of $T_{p}$ with increased values of $E_{hop}$.  The low temperature resistivity of the samples is not fit well by the model of Mercone {\it et al.}, but is better fit by the model of Jaime {\it et al.} which includes a cutoff in the single magnon scattering.  There is a strong linear correlation between the static resistivity and the magnitude of the magnon scattering term that is seen to be present in a wide range of films of various types and conditions.

\section{Acknowledgments}

The author would like to gratefully acknowledge the assistance of David Knies for the RBS measurements, as well as Victor Cestone 
and Andrew Patton in the production and characterization of the films.  Some of this work was carried out at the Naval Research Lab under funding from the Office of Naval Research.

\newpage
\begin{table}[htdp]
\caption{Results from fitting Eq.{\ref{rhofit}} for the as grown films on NdGaO$_{3}$}
\begin{center}
\begin{tabular}{|c|c|c|c|c|}
\hline
$x$&$\rho_{0}$ (m$\Omega$-cm)&$T_{q}$(K)&$\rho_{2}$ (n$\Omega$-cm/K$^{2}$)&$\rho_{5}$ (f$\Omega$-cm/K$^{5}$) \\ \hline
0.333&0.271&31.7&27.1&1.99 \\ \hline
0.247&0.498&26.5&49.7&2.61\\ \hline
0.218&0.625&26.2&64.4&3.44\\ \hline
0.20&0.323&21.3&37.7&1.64 \\ \hline
0.184&0.375&23.5&40.7&2.11 \\ \hline
0.122&0.506&9.9&49.7&3.4\\ \hline
0.097&0.477&14.4&45.9&2.82\\ \hline
\end{tabular}
\end{center}
\label{rhofitdata1}
\end{table}
\begin{table}[htdp]
\caption{Results from fitting Eq.{\ref{rhofit}} for the as grown films on SrTiO$_{3}$}
\begin{center}
\begin{tabular}{|c|c|c|c|c|}
\hline
$x$&$\rho_{0}$ (m$\Omega$-cm)&$T_{q}$(K)&$\rho_{2}$ (n$\Omega$-cm/K$^{2}$)&$\rho_{5}$ (f$\Omega$-cm/K$^{5}$) \\ \hline
0.333&0.373&27.5&41.9&6.61\\ \hline
0.245&0.666&24.6&78.7&5.56\\ \hline
0.218&0.763&19.7&88.9&6.7\\ \hline
0.20&0.862&16.8&104&7.96\\ \hline
0.184&0.574&12.9&71.2&5.26\\ \hline
\end{tabular}
\end{center}
\label{rhofitdata2}
\end{table}
\newpage

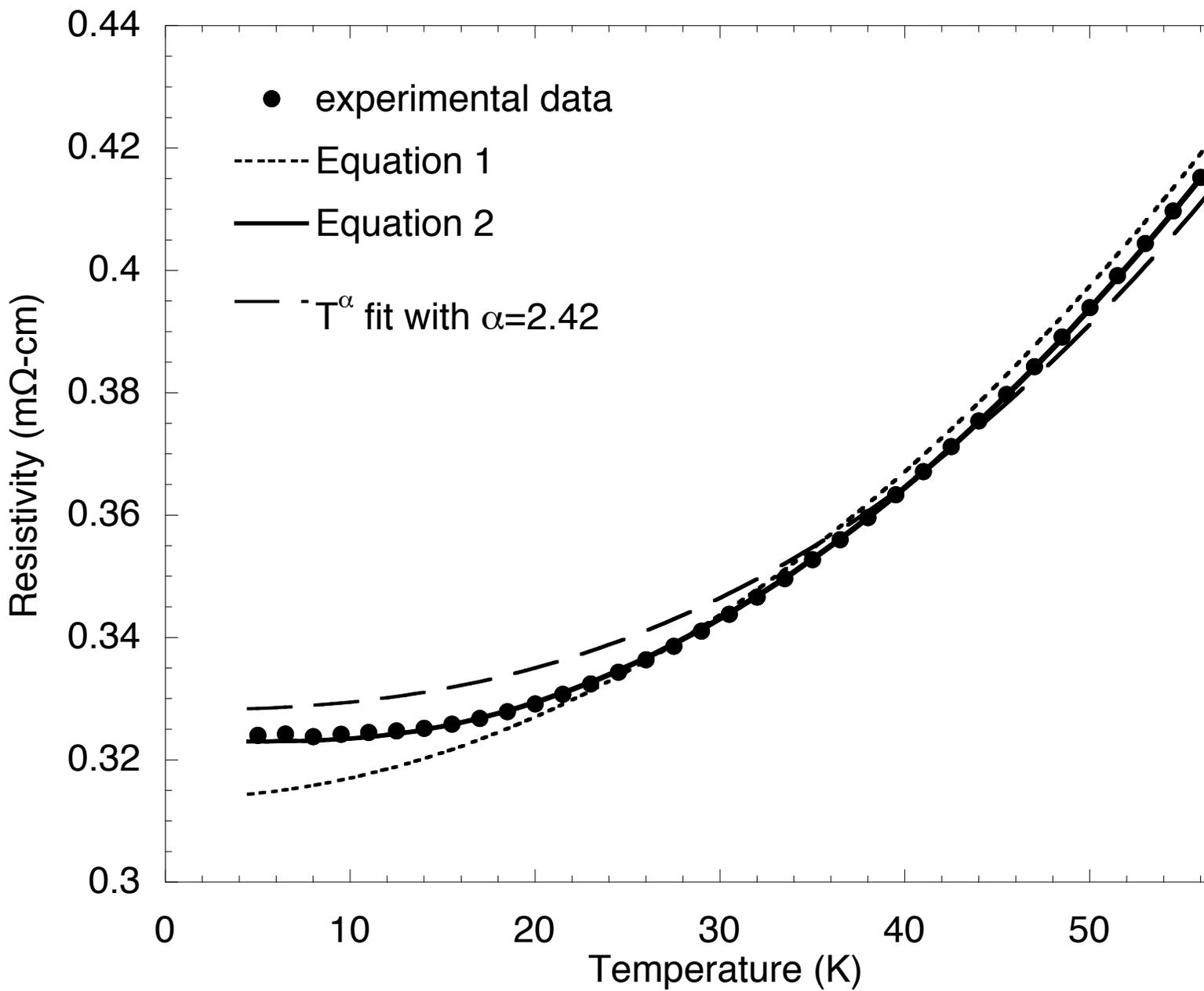
\begin{figure}
\begin{center}
\end{center}
\caption{Low temperature resistivity vs. temperature for the $x=0.2$ 
sample on NdGaO$_{3}$ along with 
the fits derived from Eq. \ref{rhofit}, Eq. \ref{rhoT} and the Mercone {\it et al.}\cite{Mercone} model}
\label{lowrho2}
\end{figure}
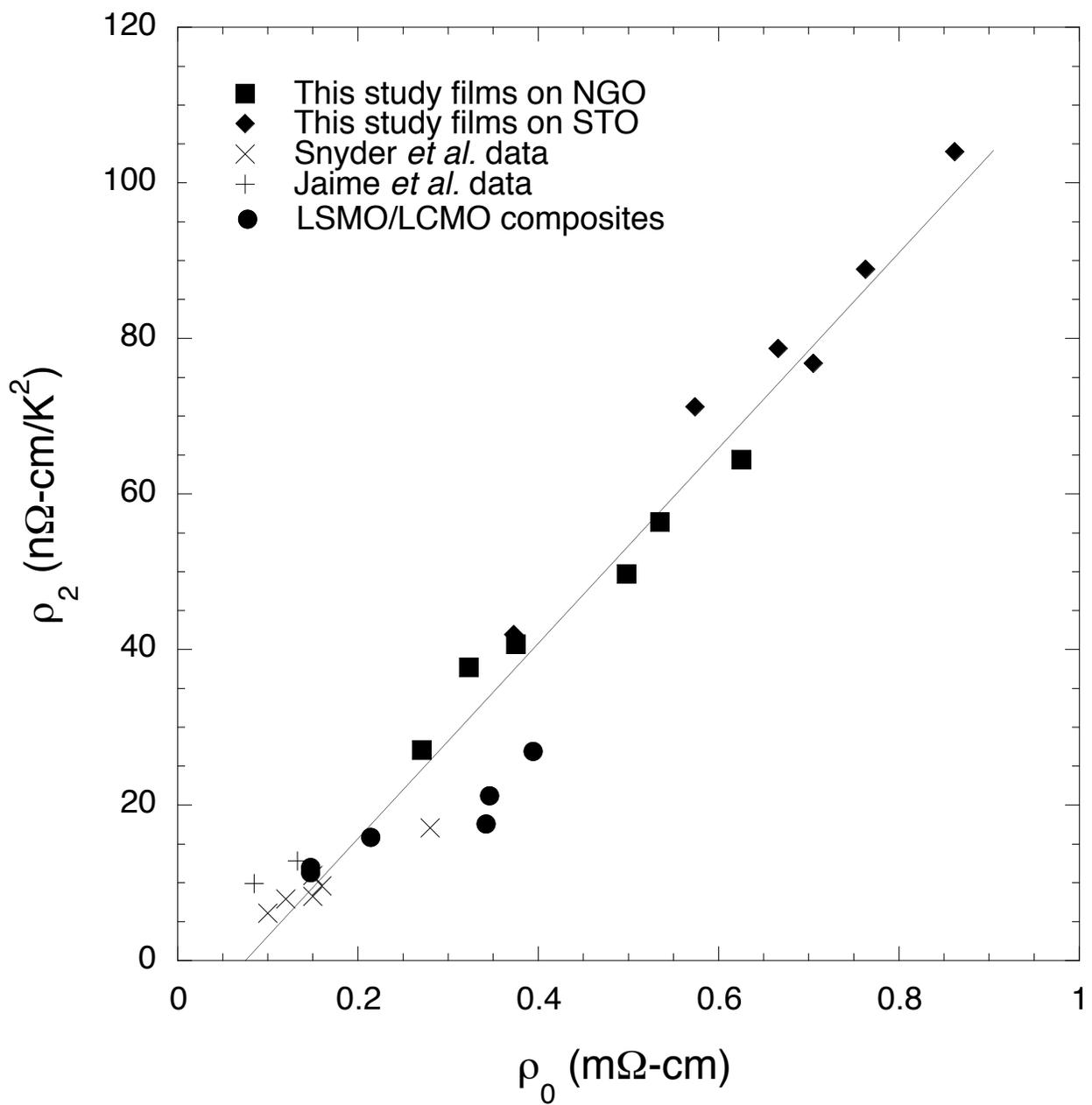
\begin{figure}
\begin{center}
\end{center}
\caption{Values of $\rho_{2}$ vs $\rho_{0}$ for the films in this work, LSMO/LCMO composite films, \cite{PRB2}, Pb-doped LCMO crystals,\cite{Jaime1} and LCMO thin 
films \cite{Snyder}.  The line is a guide to the eye.}
\label{r2r0}
\end{figure}

\newpage

\begin{thebibliography}{}

\bibitem{Helmolt}
R. von Helmolt, J. Wecker, B. Holzapfel, L. Schultz, and K. Samwer, Phys. Rev. Lett. {\bf 71}, 2331 (1993).

\bibitem{Worledge}
D. C. Worledge, L Mi\'eville, and T. H. Geballe, Phys. Rev. B{\bf 57}, 15267 (1998).

\bibitem{Song}
X. F. Song, G. J. Lian, and G. C. Xiong, Phys. Rev. B{\bf 71}, 214427 (2005).

\bibitem{Hartinger}
Ch. Hartinger, F. Mayr, A. Loidl, and T. Kopp, Phys. Rev. B{\bf 73}, 24408 (2006).

\bibitem{PRB1}
P.R. Broussard and V.C. Cestone, Mat. Res. Soc. Symp. Proc. {\bf 574}, 181 (1999).

\bibitem{Prellier}
W. Prellier, M. Rajeswari, T. Venkatesan, and R. L. Greene, Appl. Phys. Lett. {\bf 75}, 1446 (1999).

\bibitem{Mercone}
S. Mercone, C.A. Perroni, V. Cataudella, C. Adamo, M. Angeloni, C. Aruta, G. DeFilippis, F. Miletto, A. Oropallo, P. Perna, Y. Petrov, U. S. diUccio, and L. Maritato, Phys. Rev. B{\bf 71}, 64415 (2005).

\bibitem{Jaime1}
M. Jaime, P. Lin, M.B. Salamon, and P.D. Han, Phys. Rev. B {\bf 58}, R5901 (1998).

\bibitem{PRB2}
P.R. Broussard, S.B. Qadri, V.M. Browning, and V.C. Cestone, J. Appl. 
Phys. {\bf 85}, 6563 (1999).
	
\bibitem{VdP}
L. J. van der Pauw, Phillips Res.  Rep.  {\bf 13}, 1 (1958).

\bibitem{Huang}
Q. Huang, A. Santoro, J.W. Lynn, R.W. Erwin, J.A. Borchers, J.L. 
Peng, K. Ghosh, and R.L. Greene, Phys. Rev. B {\bf 58}, 2684 (1998).

\bibitem{Schiffer}
P. Schiffer, A.P. Ramirez, W. Bao, and S.-W. Cheong, Phys. Rev. Lett. {\bf 75}, 3336 (1995).

\bibitem{Browning}
V.M. Browning, R.M. Stroud, W.W. Fuller-Mora, J.M. Byers, M.S. 
Osofsky, D.L. Knies, K.S. Grabowski, D. Koller, J. Kim, D.B. Chrisey, 
and J.S. Horwitz, J. Appl. Phys.  {\bf 83}, 7070 (1998).

\bibitem{WZ}
X. Wang and X.-G. Zhang, Phys. Rev. Lett. {\bf 82}, 4276 (1999).

\bibitem{NelMed}
J.A. Nelder and R. Mead, Computer Journal {\bf 7}, 308 (1965).

\bibitem{Lynn}
J.W. Lynn {\it et al.}, Phys. Rev. Lett. {\bf 76}, 4046 (1996).

\bibitem{Baca}
J.A. Fernandez-Baca, P. Dai, H.Y. Hwang, C. Kloc, and S.W. Cheong, Phys. Rev. Lett. {\bf 80}, 4012 (1998).

\bibitem{Snyder}
G.J. Snyder, R. Hiskes, S. DiCarolis, M.R. Beasley, and T.H. Geballe, 
Phys. Rev. B. {\bf 53}, 14434 (1996).

\end{thebibliography}
\end{document}